\documentclass{article}
\pdfoutput=1
\usepackage[utf8]{inputenc}
\usepackage[normalem]{ulem}
\usepackage{tikz,subcaption}
\usepackage{pgf}
\usetikzlibrary{decorations.pathreplacing}
\usetikzlibrary{arrows,shapes}
\usepackage{amsfonts,mathtools,multicol,amssymb}
\usepackage{fullpage}
\usepackage{jcappub}

\newcommand{\fnl}{f_{\text{NL}}}

\newcommand{\tth}{\tfrac{3}{2}}
\newcommand{\rhoh}{\left( \frac{\rho}{H} \right)}

\title{Influence of Super-Horizon Modes on Correlation Functions during Inflation}
\author{Anne-Sylvie Deutsch}\emailAdd{afd126@psu.edu}
\affiliation{Institute for Gravitation and the Cosmos \& Physics Department, The Pennsylvania State University, University Park, PA 16802, USA}
\preprint{IGC-17/4-1}

\abstract{
Coupling between sub- and super-Hubble modes can affect the locally observed statistics of our universe. In the context of Quasi-Single Field Inflation, we can compute correlation functions and derive the influence of those unobservable modes on observed correlation functions as well as on the inferred cosmological parameters. We study how different classes of diagrams affect the bispectrum in the squeezed limit; in particular, while contact-like diagrams leave the scaling between the long and short modes unchanged, exchange-like diagrams do modify the shape of the bispectrum. We show that the mass of the hidden sector field can hence be biased by an unavoidable cosmic variance that can reach a 1-$\sigma$ uncertainty of $\mathcal{O}(10\%)$ for a weakly non-Gaussian universe. Finally, we go beyond the bispectrum and show how couplings between unobservable and observable modes can affect generic correlation functions with arbitrary order non-derivative self-interactions.
}

\begin{document}
\maketitle
\flushbottom

\tableofcontents

\section{Introduction} 

In the context of single clock inflation, statistics beyond the two-point correlation function of the primordial curvature density perturbation is a powerful tool to discriminate between different inflationary models. This is, among other things, due to the fact that mode coupling between modes of very different amplitude is negligible. Therefore, even though we might be observing only a fraction of the whole patch generated during inflation, the local, observed statistics will not be affected by larger, unobservable modes~\cite{Tanaka2011,Pajer2013}.

However, the situation changes when we leave single clock inflation~\cite{LoVerde2013,Baytas2015,Arkani-Hamed2015}, where the consistency relation no longer holds. To illustrate this, we will focus on Quasi Single Field Inflation~\cite{Chen2010a}, where an additional field with a mass $m \lesssim H$ is introduced on top of the inflaton field: 
\begin{equation}
	\mathcal{L}_{QSF} = -\frac{1}{2} (\partial \varphi)^2 + \rho \dot{\varphi} \sigma  -\frac{1}{2} (\partial \sigma)^2 - \frac{1}{2} m^2 \sigma^2 -V(\sigma),
	\label{eq:Lagrangian}
\end{equation}
where $\varphi$ is the inflaton perturbation and $\sigma$ the hidden sector field perturbation. The transfer term $\rho \dot{\varphi}\sigma$ couples the two sectors, allowing non-Gaussianities to be generated in the hidden sector, and then be transfered in the inflationary, observable sector. The field~$\sigma$ is not constrained by the approximate shift-symmetry of the inflaton, and can therefore have a large range of self-interactions $V(\sigma)$. This enables the coupling between modes of different magnitudes, while single-field models typically couple modes of similar wavelength instead.

This is not without consequence. Couplings between modes of different magnitudes implies coupling between super-horizon modes\footnote{While the horizon is not the same quantity as the Hubble length, we will refer to observable and unobservable modes as sub- and super-horizon modes respectively.} -- modes whose wavelength is larger than our Hubble, observable universe -- and sub-horizon modes -- modes living inside our Hubble volume. Therefore, these unobservable super-horizon modes can potentially induce a bias on the observed statistics in our Hubble inflationary patch, giving rise to an unavoidable cosmic variance on the locally observed cosmological parameters.

For instance, in the presence of local, scale-dependent non-Gaussianities, mode coupling does create a cosmic variance, which, depending on $\fnl$, can significantly affect the spectral index~\cite{Bramante2013}. This is not limited to the scalar sector; the presence of tensor modes can give rise to a four-point function~\cite{Dimastrogiovanni2015,Dimastrogiovanni2014}

We here focus on the cosmic variance in the squeezed limit of the bispectrum, which is the momentum configuration where one of the three external modes $k_L$ is much longer (and therefore has a small magnitude in Fourier space) than the two others $k_S$. Owing to the consistency relation~\cite{Maldacena2003}, a large signal in the squeezed limit of the bispectrum is generically considered to be a signature of multi-field inflationary models -- even if non-Bunch-Davies initial states~\cite{Chen2007,Holman2008,Agullo2011,Ganc2011,Ganc2012,Agullo2012,Ashoorioon2011,Ashoorioon2014} or the existence of a non-attractor phase~\cite{Namjoo2012,Martin2013,Chen2013} could alter the conclusion. Evidence of additional fields besides the inflaton could give us hints of the ultra-violet completion of the theory~\cite{Baumann2011,Lee2016}, and it is therefore important to understand the implications of unobservables modes on the bispectrum.

In Quasi Single Field Inflation, the squeezed bispectrum has a characteristic scaling~\cite{Chen2010a,Arkani-Hamed2015}:
\begin{equation}
	B(k_S,k_S,k_L) \propto P(k_L) P(k_S) \left(\frac{k_L}{k_S}\right)^{3/2-\nu},
	\label{eq:bisp-squeezed-limit-basic}
\end{equation}
where $P(k)$ is the power spectrum
\begin{equation}
	P(k) = \frac{\Delta_{\zeta}}{2\pi^2 k^3}
\end{equation}
and the parameter $\nu$ is
\begin{equation}
	\nu \equiv \sqrt{ \frac{9}{4}-\frac{m^2}{H^2}},
	\label{eq:definition-nu}
\end{equation}
$m$ being the mass of the additional field. Measuring the scaling $3/2-\nu$ between the long mode $k_L$ and short mode $k_S$ of the bispectrum in this limit would therefore tell us about the mass of the hidden sector field. However, owing to coupling between sub- and super-horizon modes, that scaling could be affected by the realization of the super-horizon modes, and the locally measured mass could be biased. 

In previous work~\cite{CosmicVariance2015}, we have studied the impact that mode coupling has on the squeezed bispectrum for a subset of diagrams. We found that for \emph{contact-like} diagrams, the scaling of the bispectrum in the squeezed limit is not affected by coupling between sub- and super-horizon modes. The question is now to understand how this result holds if we consider a larger set of diagrams, and is the focus of this work.

But we can go beyond the bispectrum, as we expect that super-horizon modes will also affect other correlation functions. We therefore also use similar techniques as the ones we used for the bispectrum to show how super-horizon modes can affect generic correlation functions. We will see that again, the influence of the unobservable modes will depend on the way they couple to the sub-horizon modes, either in a contact or exchange fashion.

The paper is organized as follows: In the first section, we review the framework presented in~\cite{CosmicVariance2015} to compute late-time correlators. We then generalize that previous work and derive the induced variance from super-horizon modes on $\nu$ from the infinite set of exchange-like diagrams generated by cubic self-interactions of the hidden sector field, then from arbitrary order self-interactions, and finally for derivative self-interactions. In the second section, we study how the shape and amplitude of generic correlation functions can be affected by couplings to unobservable modes.

Throughout the paper, we will use the notation $k_S$ to designate short modes, $k_L$ for long modes that are still sub-horizon, and $k_H$ or $q$ for super-horizon modes. We will denote the inflaton perturbation by~$\varphi$, and the hidden sector field perturbation by~$\sigma$.

\subsection{Late-time correlators}

Since we are aiming to understand how an arbitrary number of super-horizon modes can affect observed correlation functions, we will introduce a framework that will make these calculations more systematic and easier to carry out. The idea is to compute late-time correlators using the expansion of the curvature perturbation in terms of a Gaussian field. Each term of the series will contain the particle physics corresponding to the influence of the super-horizon modes on the corresponding correlation function. We use correlation functions of the curvature perturbation $\zeta$ rather than the inflaton field $\varphi$ since this corresponds to the observable quantity that is eventually measured, and one can always work in a specific gauge in which, at first order, $\varphi$ and $\zeta$ are linearly related to each other:
\begin{equation}
	\varphi = -\sqrt{2\epsilon} \zeta,
\end{equation}
$\epsilon$ being one of the slow-roll parameters. It is therefore easy to express one correlation function in terms of the other.

Correlation function are defined by:
\begin{equation}
	\langle \zeta_{\text{NG}} (\mathbf{k}_1) \dots \zeta_{\text{NG}} (\mathbf{k}_n) \rangle \equiv (2\pi)^3 \, \delta^{(3)}(\mathbf{k}_1 + \dots + \mathbf{k}_n) \, F_n(\mathbf{k}_1,\dots,\mathbf{k}_n),
\end{equation}
where $\zeta_{\text{NG}}(\mathbf{k})$ is the curvature perturbation in Fourier space. We are going to compute these correlation functions in a systematic way by developing $\zeta_{\text{NG}}(\mathbf{k})$ in a series around a Gaussian field $\zeta(\mathbf{k})$
\begin{equation}
	\zeta_{\text{NG}}(\mathbf{k}) = \zeta(\mathbf{k}) + \fnl Z_2 (\mathbf{k}) + g_{\text{NL}} Z_3 (\mathbf{k}) + \dots.
	\label{eq:expansion-zeta-fourier-space}
\end{equation}
The $n$-th order term of the sum is given by:
\begin{equation}
\begin{split}
	Z_n(\mathbf{k}) & = \frac{(2\pi)^3}{n!} \prod_{\ell = 1}^n \int \frac{d^3p_{\ell}}{(2\pi)^3} \left[ \zeta(\mathbf{p}_1) \dots \zeta(\mathbf{p}_n) - \mathcal{F}(\zeta(\mathbf{p}_1), \dots, \zeta(\mathbf{p}_n)) \right] \\
	& \qquad \times N_n(\mathbf{p}_1, \dots,\mathbf{p_{n}}, \mathbf{k}) \ \delta^{(3)} (\mathbf{k}- \mathbf{p}_1 - \, \dots \, - \mathbf{p}_n)
\end{split}
\label{eq:nth-order-term-expansion}
\end{equation}
where $\mathcal{F}(\zeta(\mathbf{p}_1), \dots, \zeta(\mathbf{p}_n))$ 
ensures that the mean of $\zeta_{\text{NG}}(\mathbf{k})$ is zero, and that we only get contributions to the connected parts of the correlation function. The kernel $N_n(\mathbf{p}_1,\dots,\mathbf{p}_n,\mathbf{k})$ is symmetric in its $n$ first entries $p_i$, and is chosen to reproduce the tree-level $(n+1)-$point function; therefore, it will have a dependence on the parameters of the Lagrangian.

Since we are interested in studying the influence of super-horizon modes on correlation functions as observed in a sub-volume, we perform a split between super-horizon modes $k_H$, whose associated wavelengths are bigger than the Hubble scale $k_H < k_0 \simeq 2\pi / H$, and sub-horizon modes $k_S \geq k_0$ that are inside the observable volume. Hence, integrals can be written as:
\begin{equation}
	\int \frac{d^3k}{(2\pi)^3} \quad \to \quad \int_{k<k_0} \frac{d^3k}{(2\pi)^3} + \int_{k\geq k_0} \frac{d^3k}{(2\pi)^3}.
\end{equation}
Expanding each integral in the $n-$th order term~\eqref{eq:nth-order-term-expansion} of the series gives rise to $n$ terms\footnote{The $Z_n^i(\mathbf{k})$ terms must contain at least one observable mode. We hence ignore any terms involving only integrals over super-horizon modes, with no integral over sub-horizon modes.}:
\begin{equation}
	Z_n(\mathbf{k}) = Z_n^{(0)}(\mathbf{k}) + \dots + Z_n^{(n-1)}(\mathbf{k}),
\end{equation}
where $Z_n^{(i)}(\mathbf{k})$ corresponds to the term with $i$ integrals over super-horizon modes:
\begin{equation}
\begin{split}
	Z_n^{(i)}(\mathbf{k}) & = \frac{(2\pi)^3}{n!} \prod_{\ell = 1}^{n-i} \left( \int_{p_{\ell}\geq k_0} \frac{d^3p_{\ell}}{(2\pi)^3} \right) \prod_{\ell = n-i+1}^n \left( \int_{p_{\ell} < k_0} \frac{d^3p_{\ell}}{(2\pi)^3} \right) N_n(\mathbf{p}_1, \dots,\mathbf{p_{n}}, \mathbf{k}) \\
	& \qquad \times \delta^{(3)} (\mathbf{k}- \mathbf{p}_1 - \, \dots \, - \mathbf{p}_n) \left[ \zeta(\mathbf{p}_1) \dots \zeta(\mathbf{p}_n) - \mathcal{F}(\zeta(\mathbf{p}_1), \dots, \zeta(\mathbf{p}_n)) \right].
\end{split}
\end{equation}

Rearranging each term of~\eqref{eq:expansion-zeta-fourier-space} according to the number of sub-horizon modes involved in each integral, one obtains the field as observed in the sub-volume:
\begin{align}
\begin{split}
	\zeta_{\text{NG}}^{\text{obs}} (\mathbf{k}) & = \left[ \zeta(\mathbf{k}) + 2 \fnl \, Z_2^{(1)}(\mathbf{k}) + 3 g_{\text{NL}} \, Z_3^{(2)}(\mathbf{k}) + 4 h_{\text{NL}} \, Z_{4}^{(3)} (\mathbf{k}) + \dots \right] \\
	& \quad + \left[ \fnl \, Z_2^{(0)}(\mathbf{k}) + 3 g_{\text{NL}} \, Z_3^{(1)}(\mathbf{k}) + 6 h_{\text{NL}} Z_4^{(2)}(\mathbf{k}) + \dots \right] \\
	& \quad + \left[ g_{\text{NL}} \, Z_3^{(0)}(\mathbf{k}) + 4 h_{\text{NL}} \, Z_4^{(1)}(\mathbf{k}) +  \dots \right] + \dots
\end{split} \label{eq:expansion-zng-after-split} \\
& \equiv Z_1^{\text{obs}}(\mathbf{k}) + \fnl^{\text{obs}} Z_2^{\text{obs}}(\mathbf{k}) + g_{\text{NL}}^{\text{obs}} Z_3^{\text{obs}}(\mathbf{k}) + \dots
\end{align}
where numerical pre-factors are due to the symmetry of $Z_n(\mathbf{k})$ in the integrated over momenta $p_i$. The first line of~\eqref{eq:expansion-zng-after-split} corresponds to the linear term of the expansion of the field $\zeta(\mathbf{k})$ as observed in the sub-volume; in other words, it contains all the terms of the expansion involving only one field with a sub-horizon mode. The next lines are the next order terms gathering all the terms with two, three, etc sub-horizon modes.

This allows us to write each term with a local, effective kernel $N_n^{\text{obs}}(\mathbf{p}_1, \dots, \mathbf{p}_n,\mathbf{k})$:
\begin{equation}
\begin{split}
	c_{\text{NL}}^{(n)\text{obs}} Z_n^{\text{obs}}(\mathbf{k}) & = \frac{c_{\text{NL}}^{(n)}}{n!} \prod_{i=1}^n \left( \int_{p_i>k_0} \frac{d^3p_i}{(2\pi)^3} \right) [\zeta(\mathbf{p}_1) \dots \zeta(\mathbf{p}_n)- \mathcal{F}(\zeta(\mathbf{p}_1),\dots,\zeta(\mathbf{p}_n))] \\
	& \qquad \times N_n^{\text{obs}}(\mathbf{p}_1,\dots,\mathbf{p}_n,\mathbf{k}) \ (2\pi)^3 \delta^{(3)}(\mathbf{k}-\mathbf{p}_1 - \dots - \mathbf{p}_n)
\end{split}
\end{equation}
where $c_{\text{NL}}^{(2)} = \fnl$, $c_{\text{NL}}^{(3)}=g_{\text{NL}}$, etc, and
\begin{equation}
\begin{split}
	N_n^{\text{obs}} (\mathbf{p}_1, \dots, \mathbf{p}_n, \mathbf{k}) & = N_n(\mathbf{p}_1, \dots, \mathbf{p}_n, \mathbf{k}) + \sum_{i = n+1} \frac{i!}{n!(i-n)!} \frac{c_{\text{NL}}^{(i)}}{c_{\text{NL}}^{(n)}}\\
	& \qquad \times \prod_{j=n+1}^i \left( \int_{p_j<k_0} \frac{d^3p_j}{(2\pi)^3} \zeta(\mathbf{p}_j) \right) N_i(\mathbf{p}_1, \dots, \mathbf{p}_i, \mathbf{k}).
\end{split}
\end{equation}
We see that the observed kernel gets correction from higher-order kernels involving more and more super-horizon modes.

\section{The Bispectrum in the Squeezed Limit}
We can now apply the framework presented above to a particular case: the computation of the bispectrum in the squeezed limit. This will allow us to understand how super-horizon modes can affect the $3/2-\nu$ scaling of the function, and therefore how protected is the measurement of the mass of the hidden sector.

\begin{figure}
	\centering \includegraphics{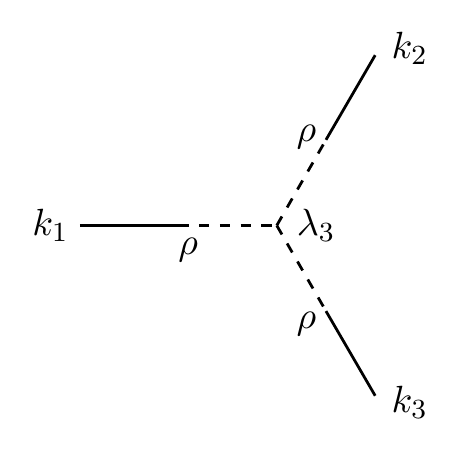}
	\caption{The mean, tree-level bispectrum in Quasi-Single Field Inflation. The solid lines represent the inflaton fields, while the dashed lines correspond to the hidden sector field. Transfer vertices connect the inflaton to the hidden field, and the hidden field is self-interacting.}
	\label{fig:basic-bispectrum}
\end{figure}

In order to apply this framework, we will need to evaluate correlation functions in order to determine their scaling with respect to the different involved scales. We hence quickly illustrate how to estimate the correlation function in one simple case: the squeezed limit of the bispectrum in the large volume, i.e. not including yet effects related to unobservable modes. The Lagrangian is given by~Eq.\eqref{eq:Lagrangian}, and we will only assume a cubic self-interaction. The estimations follow those presented in~\cite{Baumann2011}. The correlation function can be schematically represented by:
\begin{equation}
\begin{split}
	\langle \zeta(\mathbf{k}_S) \zeta(\mathbf{k}_S) \zeta(\mathbf{k}_L)  \rangle & \sim \left. \zeta(\mathbf{k}_S) \zeta(\mathbf{k}_S) \zeta(\mathbf{k}_S) \right|_{\tau=0} \prod_{i=1}^3 \int d\tau_i \, a^3 \rho \varphi^{\prime}(\mathbf{k}_i) \sigma (\mathbf{k}_i) \\
	& \qquad \times \int d\tau a^4 \lambda_3 \sigma(\mathbf{k}_S) \sigma(\mathbf{k}_S) \sigma(\mathbf{k}_L),
\end{split}
\end{equation}
where the prime is a time derivative with respect to conformal time $\tau$. While this expression is not complete, its structure reveals the most important features that we are looking for. The three first integrals correspond to the three transfer vertices between the inflaton field and the hidden sector field, see Fig.~\ref{fig:basic-bispectrum}. The last integral corresponds to the central cubic vertex.

Each $\zeta(\mathbf{k})$ evaluated at $\tau=0$ gives a factor of $\Delta_{\zeta}/k^{3/2}$. To compute the integrals, we replace each field by its corresponding mode function:
\begin{align}
	\text{Massless field, } \varphi &  & u_k(\tau) & = \frac{H}{\sqrt{2k^3}} (1+ik\tau) e^{-ik\tau}, \\
	\text{Massive field, } \sigma & & v_k(\tau) & = \frac{H\sqrt{\pi}}{2\sqrt{k^3}} (-k\tau)^{3/2} H_{\nu}^{(1)} (-k\tau) \nonumber \\
	\text{When } |k\tau |\to 0 & & & \sim -i \frac{2^{\nu}\Gamma(\nu)}{2\sqrt{\pi}} \frac{H}{\sqrt{k^3}} (-k\tau)^{3/2-\nu}.
\end{align}
For the first three integrals, the exponential of the massless mode function will suppress the integrand when $|k\tau|\gg1$. The integral will then dominated for $\tau\sim -k^{-1}$:
\begin{align}
	\int d\tau a^3  \varphi^{\prime}(\mathbf{k}) \sigma(\mathbf{k}) \ \to \ \int d\tau a^3  u_k^{\prime}(\tau) v_k(\tau) & \sim \int d\tau \frac{\rho}{(-H\tau)^3} \frac{H^2}{\sqrt{2k^3}} \frac{k^2 \tau}{k^{\nu}} (-\tau)^{\tfrac{3}{2}-\nu} e^{-ik\tau} \nonumber \\
	& \sim \frac{\rho}{H} k^{\tfrac{1}{2}-\nu} \int d \tau \ \tau^{-\tfrac{1}{2}-\nu} e^{-ik\tau}\nonumber \\
	& \sim \frac{\rho}{H}
\end{align}
For the cubic vertex integral, the mode dominating the integral is the shortest one as it is the last mode to cross the horizon, so $\tau\sim -k_S^{-1}$
\begin{align}
	\int d\tau a^4 \lambda_3 \, \sigma(\mathbf{k}_S) \sigma(\mathbf{k}_S) \sigma(\mathbf{k}_L) \ \to \ \int d\tau a^4 \lambda_3 \, v^2_{k_S}(\tau) \, v_{k_L}(\tau) & \sim \int d\tau \frac{\lambda_3}{(-H\tau)^4} \frac{H^3}{k_S^{2\nu} k_L^{\nu}} \tau^{3(\tfrac{3}{2}-\nu)} \nonumber \\
	& \sim \frac{\lambda_3}{H} k_L^{-\nu} k_S^{-2\nu} \int d\tau \ \tau^{\frac{1}{2}-3\nu} \nonumber \\
	& \sim \frac{\lambda_3}{H} k_S^{\nu-\tfrac{3}{2}} k_L^{-\nu}.
\end{align}
Then, combining everything, we obtain:
\begin{align}
	\langle \zeta(\mathbf{k}_S) \zeta(\mathbf{k}_S) \zeta(\mathbf{k}_L) \rangle & \sim \frac{\Delta_{\zeta}^3}{k_S^3k_L^{3/2}} \left(\frac{\rho}{H}\right)^3 \frac{\lambda_3}{H} k_S^{\nu-\tfrac{3}{2}} k_L^{-\nu} \nonumber \\
	& \sim  \frac{1}{\Delta_{\zeta}} \left(\frac{\rho}{H} \right)^3 \frac{\lambda_3}{H} P(k_L) P(k_S) \left(\frac{k_L}{k_S} \right)^{\tfrac{3}{2}-\nu},
\end{align}
which corresponds to the result Eq.~\eqref{eq:bisp-squeezed-limit-basic}.

\subsection{Exchange Diagrams with Cubic Couplings}
\label{sec:understanding-exchange-diagrams}

\begin{figure}[htbp]
	\centering
	\includegraphics[width=0.9\textwidth]{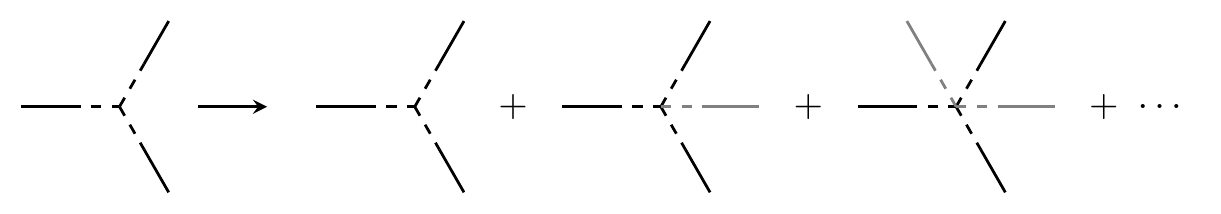}
	\caption{Subset of diagrams included in the previous analysis of the scaling of the bispectrum in the squeezed limit~\cite{CosmicVariance2015}. All super-horizon modes were only attached to the central vertex, and not to sub-horizon modes.}
	\label{fig:previous-work}
\end{figure}

\begin{figure}
	\centering \includegraphics[scale=1]{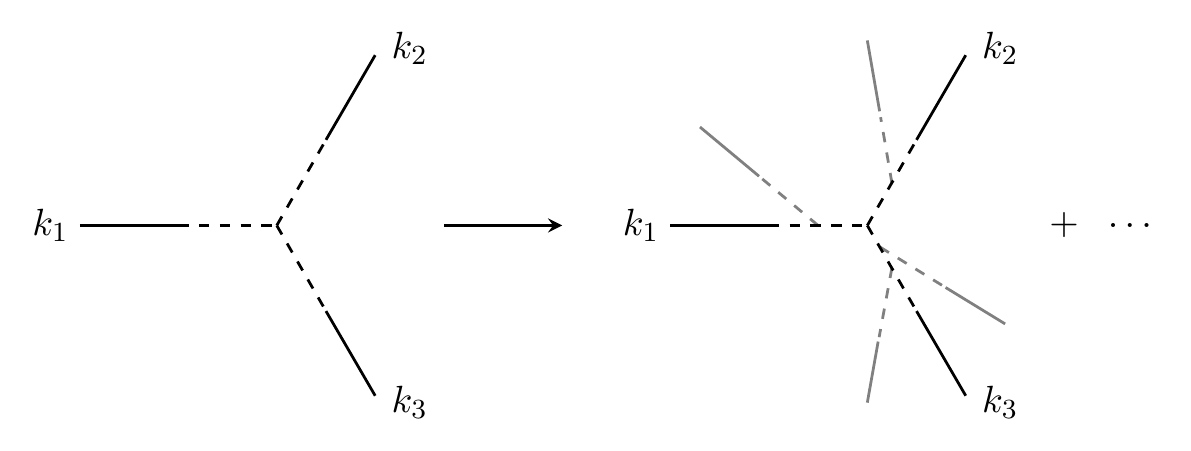}
	\caption{To evaluate the correlation function with $N$ long modes, we take the initial correlation function that we want to study (in black, on the left), and add super-horizon modes (in gray, on the right). Each super-horizon mode will be attached to one of the initial momenta $k_1, k_2$ or $k_3$. The initial correlation function acts as a backbone onto which the super-horizon modes are connected.}
	\label{fig:add-long-modes-cubic}
\end{figure}

In~\cite{CosmicVariance2015}, it was shown that the scaling of the ratio of the long and short modes of the bispectrum in the squeezed limit is preserved from super-horizon mode contributions if all the super-horizon modes are connected to the same vertex as the sub-horizon modes -- or, in other words, for contact-like diagrams, see Fig.~\ref{fig:previous-work}. We now investigate how the coupling is affected if one allows the super-horizon modes to connect to other parts of the diagram, in an exchange-like fashion, as shown on Fig.~\ref{fig:add-long-modes-cubic}. We will first only consider cubic self-interactions of the hidden sector field:
\begin{equation}
	V(\sigma) = \frac{\lambda_3}{3!} \sigma^3,
\end{equation}
and generalize the result to arbitrary order self-interactions in the next sub-section~\ref{sec:generalization-to-arbitrary-order-non-derivative-couplings}.

In order to derive correlation functions for exchange-like diagrams using the late-time correlators formalism, the first step is to derive the kernels. The following expression is an approximation of the full result that gives the usual bispectrum in the squeezed limit~\cite{Chen2010a}:
\begin{equation}
	N_2(p_1,p_2,k) \propto \frac{(p_1+p_2+k)^{3\nu-\tfrac{3}{2}}}{(p_1 p_2 k)^{\tfrac{3}{2}+\nu}} p_1^3 p_2^3.
\end{equation}
This kernel has no super-horizon modes. The corresponding kernel with $N$ super-horizon modes will be denoted by the function $N_{N+2} (p_1,p_2,k,q_1,\dots,q_N)$. In the following, a mode $q_i$ will in the end be taken to be a super-horizon mode, so we directly denote them with a different letter in order to make the notations more transparent.

To derive the $N+3$-point correlation function $F_{N+3}(p_1,p_2,k,q_1,\dots,q_N)$, we first notice that:
\begin{itemize}
\item Adding one long mode to a diagram means adding one external leg, one transfer vertex and one cubic self-coupling.
\item As the super-horizon mode's momentum $q$ is much smaller than the sub-horizon modes, it is not going to affect the momentum $k$ of the leg upon which it is connected, and so $k + q \simeq k$.
\end{itemize}
Therefore, we can associate each long mode to one of the three sub-horizon modes onto which it connects. For a total of $N=n+m+\ell$ super-horizon modes, $n$, $m$ and $\ell$ will be attached to the external leg with momentum $p_1$, $p_2$ and $k$ respectively.

Therefore, we want to evaluate the following integrals:
\begin{equation}
\begin{split}
	F_{N+3} (p_1,p_2,k,q_1,\dots,q_N) \propto & \left( \frac{\rho}{H} \right)^{N+3} \hspace{-1em} \frac{\Delta^{N+3}}{(p_1 p_2 k q_1 \dots q_N)^{3/2}} \left[ \int d \tau a^4 \lambda_3 \sigma(p_1) \sigma(p_1) \sigma(q) \right]^n   \\
	&  \left[ \int d \tau a^4 \lambda_3 \sigma(p_2) \sigma(p_2) \sigma(q) \right]^m \left[ \int d \tau a^4 \lambda_3 \sigma(k) \sigma(k) \sigma(q) \right]^{\ell} \\
	& \left[ \int d \tau a^4 \lambda_3 \sigma(p_1) \sigma(p_2) \sigma(k) \right],
	\label{eq:correlation-fct-N3}
\end{split}
\end{equation}
where we took $q_1 \sim \dots \sim q_N \sim q$. Each integral is evaluated in the same way as described above:
\begin{equation}
	\int d \tau a^4 \lambda_3 \, \sigma(p) \sigma(p) \sigma(q) \sim \frac{\lambda_3}{H} q^{-\nu} p^{\nu-3/2},
\end{equation}
where $p$ refers to either $p_1, p_2$ or $k$.
In the last integral of~Eq.\eqref{eq:correlation-fct-N3} however, it is not obvious which mode is the shortest one, and therefore we choose to evaluate the integral in the following way:
\begin{equation}
	\int d \tau a^4 \lambda_3 \sigma(p_1) \sigma(p_2) \sigma(k) \sim \frac{\lambda_3}{H} (p_1 p_2 k)^{-\nu} (p_1 + p_2 + k)^{3\nu-3/2}.
\end{equation}
Combining everything together and including a combinatorics factor to take into account the number of arrangement of $N$ long modes in three sets of modes attached to either $p_1$, $p_2$ or $k$, this yields:
\begin{equation}
\begin{split}
	F_{N+3} (p_1,p_2,k,q_1,\dots,q_N) & \propto \frac{(N+2)!}{N!2!} \left(\frac{\rho \Delta}{H}\right)^{N+3} \left(\frac{\lambda_3}{H}\right)^{N+1} \\
	& \qquad \times \frac{(p_1 + p_2 +k)^{3\nu - \tfrac{3}{2}}}{(p_1 p_2 k q^N)^{\nu+\tfrac{3}{2}}} \sum_{\substack{\left\{n,m,\ell \, | \right. \\ \left. n+m+\ell=N\right\}}}  \left( p_1^{n} p_2^m k^{\ell} \right)^{\nu-3/2},
	\label{eq:FN3-cubic}
\end{split}
\end{equation}
where the sum is over partitions of $N$ in three sets (i.e. over arrangements of $n,m,\ell$ such that $n+m+\ell=N$, where $n, m$ and $\ell$ denote the number of super-horizon modes connecting to $p_1, p_2$ and $k$ respectively).

We then find the following expression for the corresponding kernel\footnote{In our notation, the correlation function $F_n$ is related to the kernel $N_{n-1}$.}:
\begin{equation}
\begin{split}
	N_{N+2} (p_1,p_2,k,q_1,\dots,q_N)& \propto (2\pi^2)^{N+2} \frac{(N+2)!}{N!2!} \left(\frac{\rho}{H}\right)^{N+3} \left(\frac{\lambda_3}{H\Delta_{\zeta}}\right)^{N+1} \\
	& \qquad \times  \frac{ (p_1 + p_2 +k)^{3\nu - \tfrac{3}{2}}}{(p_1 p_2 q^N)^{\nu-\tfrac{3}{2}}k^{\nu+\tfrac{3}{2}}} \sum_{\substack{\left\{n,m,\ell \, | \right. \\ \left. n+m+\ell=N\right\}}}\left( p_1^{n} p_2^m k^{\ell} \right)^{\nu-3/2} .
\end{split}
\end{equation}
It is then just a matter of manipulating the above expressions to derive the observed kernel in the sub-volume corrected by super-horizon modes couplings:
\begin{equation}
\begin{split}
	N_2^{\text{obs}} (p_1,p_2,k) \propto & \left(\frac{\lambda_3 \rho^3 (2\pi^2)^2}{\Delta_{\zeta} H^4} \right) \frac{(p_1+p_2+k)^{3\nu - \tfrac{3}{2}}}{(p_1 p_2)^{\nu-\tfrac{3}{2}} k^{\tfrac{3}{2}+\nu}} \sum_{N=0} \frac{(N+2)!}{N!2!} \frac{c_{\text{NL}}^{(N+2)}}{f_{\text{NL}}} \left(\frac{\lambda_3 \rho 2\pi^2}{H^2 \Delta_{\zeta}}\right)^N\\ 
	&  \times  \prod_{i=3}^{N+2} \left[ \int_{q_i<k_0} \frac{d^3q_i}{(2\pi)^3} \zeta(q_i) q_i^{\tfrac{3}{2}-\nu} \right] \sum_{\substack{\left\{n,m,\ell \, | \right. \\ \left. n+m+\ell=N\right\}}} (p_1^n p_2^m k^{\ell})^{\nu-\tfrac{3}{2}} .
\end{split}
\end{equation}
This was the last step before computing the three-point correlation function in the squeezed limit:
\begin{align}
\begin{split}
 \langle \zeta (k_1) \zeta (k_2) Z_2^{\text{obs}} (k_3) \rangle & = \frac{f_{\text{NL}}}{f_{\text{NL}}^{\text{obs}}} \frac{1}{2! (2\pi)^3} \int_{k_0} d^3p_1 d^3p_2 \left[ \langle \zeta(k_1) \zeta(k_2) \zeta(p_1) \zeta(p_2) \rangle \right. \\
 & \quad \left. - \langle \zeta (k_1) \zeta(k_2)\rangle \langle \zeta(p_1) \zeta (p_2) \rangle \right] N_2^{\text{obs}} (p_1,p_2,k) \delta^{(3)} \left( \mathbf{k}_3 - \mathbf{p}_1 - \mathbf{p}_2 \right)
\end{split}\\
 & \propto \Delta^3_{\zeta} \frac{(k_1+k_2+k_3)^{3\nu-\tfrac{3}{2}}}{(k_1k_2k_3)^{\tfrac{3}{2}+\nu}} \left[\sum_{N=0} \alpha_N \zeta_L^N \sum_{\substack{\left\{n,m,\ell \, | \right. \\ \left. n+m+\ell=N\right\}}} \left( \frac{k_1^nk_2^m k_3^{\ell}}{k_0^N} \right)^{\nu-\tfrac{3}{2}} \right], \label{eq:three-pt-fct-symmetric}
\end{align}
where the cumulative long wavelength background $\zeta_L$ is defined by
\begin{equation}
	\zeta_L = \int_{k_{\text{IR}}}^{k_0} \frac{d^3q}{(2\pi)^3} \left( \frac{q}{k_0}\right)^{\tfrac{3}{2}-\nu}  \zeta_{\text{G}} (q),
\end{equation}
and
\begin{equation}
	\alpha_N \equiv (2\pi^2)^N \frac{(N+2)!}{N!2!} \frac{c^{(N+2)}_{\text{NL}}}{f_{\text{NL}}} \left( \frac{\lambda_3 \rho}{H^2 \Delta_{\zeta}} \right)^N.
\end{equation}
Note that Eq.\eqref{eq:three-pt-fct-symmetric} is symmetric in $k_1,k_2,k_3$. However, if one consider the squeezed limit of this expression, i.e. if one takes one of the modes to be longer than the others -- but still sub-horizon ($q \ll k_0 \leq k_L \ll k_S$, $k_0$ being the largest observable scale) -- then
\begin{equation}
\begin{split}
	\langle \zeta (k_S) \zeta (k_S) \zeta (k_L) \rangle & \propto P(k_L) P(k_S) \left(\frac{k_L}{k_S} \right)^{\tfrac{3}{2}-\nu} \sum_{N=0} \alpha_N \zeta^N_L \left( \frac{k_S}{k_0} \right)^{N(\nu-\tfrac{3}{2})} \\
	& \qquad \times \sum_{i=0}^{N} 2^{(\nu-\tfrac{3}{2})(N-i)} (N-i+1) \left( \frac{k_L}{k_S} \right)^{i(\nu-\tfrac{3}{2})}.
\end{split}
\label{eq:bispectrum-squeezed-cubic-interactions}
\end{equation}
If we compare Eq.~\eqref{eq:bispectrum-squeezed-cubic-interactions} to Eq.~\eqref{eq:bisp-squeezed-limit-basic}, we see that the scaling of $ k_L / k_S$ is affected by the term $(k_L/k_S)^{i(\nu-3/2)}$ and is no longer protected. Super-horizon modes coupling to the bispectrum in an exchange-like fashion gives rise to a fundamental constraint, a cosmic variance on the measurement of $\nu$ -- and therefore, $m$. This is important, since it means that using the squeezed limit of the bispectrum as a proxy for the physics of hidden sector fields is not obvious or straightforward, as the influence of super-horizon modes cannot be put aside. The variance depends on the mass of the hidden field, and is more significant for lighter fields than for more massive ones.

This result should be compared to the previous result presented in~\cite{CosmicVariance2015}, where it was found that the scaling between the long and short modes of the bispectrum in the squeezed limit remains unchanged when super-horizon modes are connected to the diagram in a contact-like fashion, i.e. when these modes connect to the central vertex without creating any new vertices. Here, we have seen that for exchange-like diagrams, when super-horizon modes are connected to sub-horizon modes by creating a new vertex, the scaling is modified.

We now want to give a quantitative estimate of the variance on the mass of the hidden sector field. We can do that by evaluating Eq.\eqref{eq:bispectrum-squeezed-cubic-interactions} over a range of $k_L/k_S$ from $10^{-4}$ to $10^{-2}$, finding the best fit to that curve, and converting it to the corresponding observed mass, which can then be compared to the actual value of the mass in the large volume $m$. 

However, the bispectrum~\eqref{eq:bispectrum-squeezed-cubic-interactions} is not always well behaved, as the series is not always converging. In particular, the scaling of the bispectrum is very sensitive to the parameter~$\alpha_N$. We therefore have to restrict our Lagrangian parameter space to ensure that the regime is weakly non-Gaussian. This constrains the parameter $\alpha_N$, that we chose to be $\sqrt[N]{\frac{c_{\text{NL}}^{(N+2)}}{\fnl}} \frac{\lambda_3 \rho}{H^2 \Delta_{\zeta}} \sim 5 $.

An estimation for the value of the cumulative background $\zeta_L$ is given by~\cite{CosmicVariance2015}:
\begin{equation}
	\langle\zeta^2_{L}(x)\rangle=\frac{\Delta^2_{\zeta}(k_0) [1-e^{-N_{\rm extra}(2 - 2\nu +n_s)}]}{2-2\nu+n_s}\;,
\label{eq:variance-of-long-mode}
\end{equation}
where $N_{\rm extra}\equiv \ln(k_0/k_{\text{IR}})$ is the number of e-folds of inflation before the mode $k_0$ exited the horizon (the ``extra'' e-folds if $k_0$ is taken to be the largest observable scale) and $k_{\text{IR}}$ is an infrared cut-off, $k_{\text{IR}} \sim 2 \pi V_L^{1/3}$, where $V_L$ corresponds to the large volume.
We assume $N_{\text{extra}}=110$, and current Planck measurements for $\Delta_{\zeta}$ and $n_s$~\cite{PlanckCollaboration2015}.

\begin{figure}
\centering%
\begin{subfigure}{.5\textwidth}
	\centering%
	\includegraphics[scale=.9]{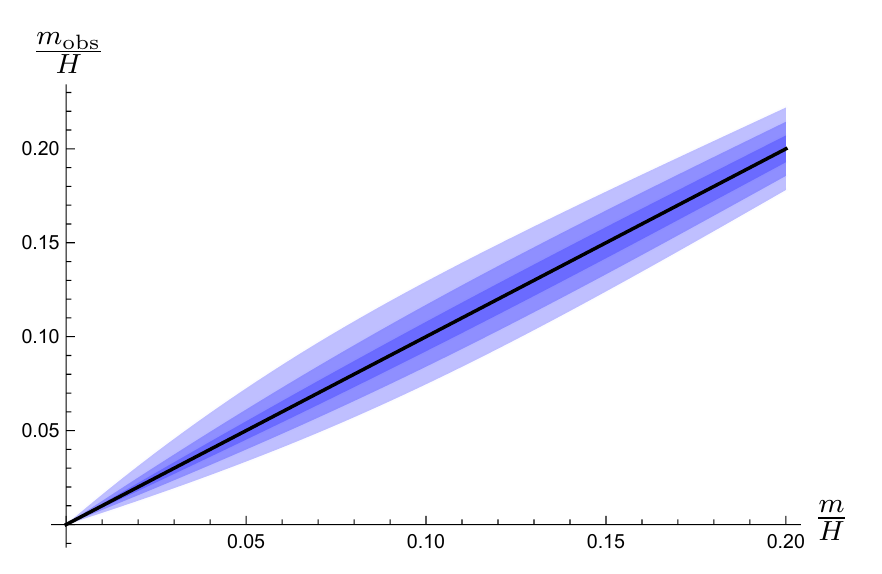}
	\caption{Contours of $m_{\text{obs}}$ w.r.t. $m$.}
	\label{fig:error-on-mass}
\end{subfigure}%
\begin{subfigure}{0.5\textwidth}
	\centering%
	\includegraphics[scale=.8]{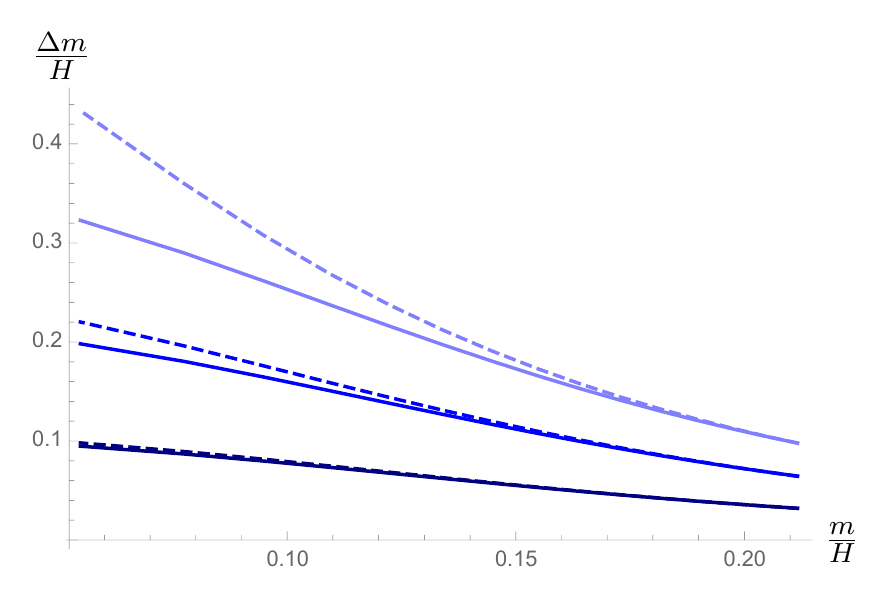}
	\caption{Relative shift on the mass.}
	\label{fig:rel-error-on-mass}
\end{subfigure}
\caption{Cosmic variance on the observed mass with respect to the actual mass, with $N_{\text{extra}}=110$. Dark, normal and light blue color corresponds to 1, 2 and 3-$\sigma$ uncertainty respectively for both figures. On the left, a plot showing the contour lines of the region where the mass could be measured in a sub-volume. The black center line corresponds to the mass of the field in the large volume. On the left, the relative shift as a function of the mass. The dashed/plain line corresponds to the variance found in a sub-volume located in an under-dense/over-dense region of the large volume, respectively. Light fields have a relatively higher cosmic variance than heavier fields, and the typical 1-$\sigma$ relative uncertainty on the mass is of $\mathcal{O}(5-10\%)$. We expect that a relaxation of the restrictions on the parameter space would lead to an increase in the cosmic variance for all masses.}
\label{fig:plots-error-on-mass}
\end{figure}

The results are presented in Fig.~\ref{fig:plots-error-on-mass}. Fig.~\ref{fig:error-on-mass} shows the contours of the observed mass in the sub-volume, with 1, 2 and 3-$\sigma$ uncertainty, while the computed relative shift $\left|\tfrac{m-m_{\text{fit}}}{m}\right| = \frac{\Delta m}{m}$ is shown in Fig.~\ref{fig:rel-error-on-mass}. The uncertainty at 1-$\sigma$ on $\nu$ is of $\mathcal{O}(10^{-3})-\mathcal{O}(10^{-4})$, which translate into a relative uncertainty on the mass of the hidden sector field of order 5-10\%. Light fields are typically more affected by the super-horizon modes than heavier fields. We expect that those uncertainties would increase if the restrictions on $\rho$ and $\lambda_3$ that we had to take are weakened, as the uncertainty on the mass is linked to the restrictions on $\alpha_N$.

Forecasts of the accuracy of $\nu$ of future galaxy surveys can be compared to the above results. Using Eq.(38) of~\cite{Sefusatti2012a}, we compute $\Delta \nu \sim .3$ for $f_{\text{NL}}=5$ and $\nu=1.5$ at 1-$\sigma$ uncertainty.

Therefore, for the selected region of the parameter space where the series remains perturbative, while the scaling of the bispectrum in the squeezed limit is affected by long modes, the effect is not strong enough to be a limiting factor for current or future measurements of the mass of the hidden sector field.

\subsection{Generalization to Arbitrary Order Non-Derivative Couplings}
\label{sec:generalization-to-arbitrary-order-non-derivative-couplings}

\begin{figure}
	\centering \includegraphics{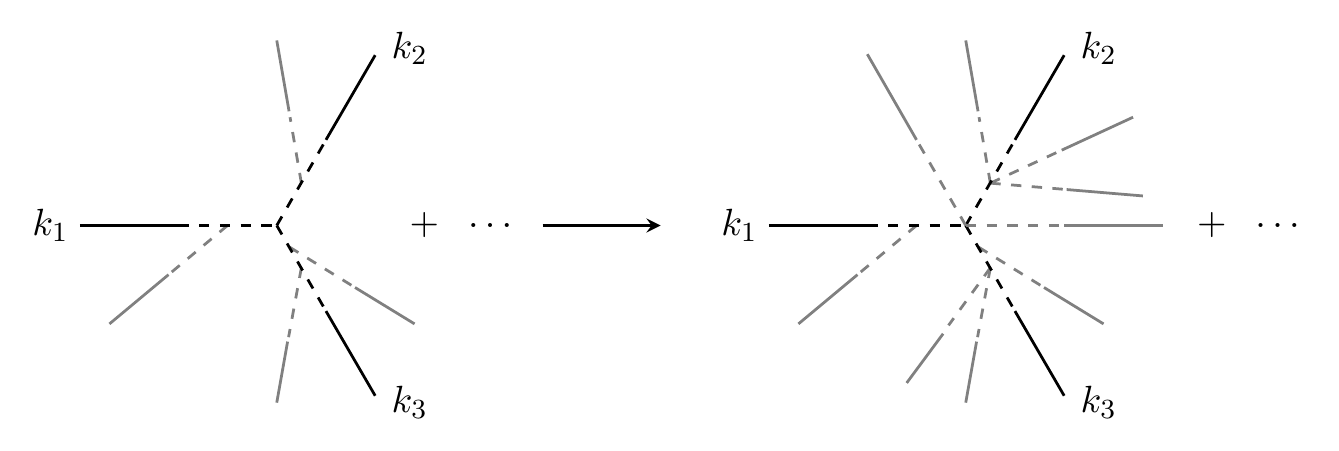}
\caption{Generalization of section~\ref{sec:understanding-exchange-diagrams}. Rather than just considering cubic coupling, we compute diagrams involving all higher-order coupling, which means that super-horizon modes can connect to any sub-horizon modes, creating vertices with an arbitrary number of edges, or to the central vertex.}
\label{fig:from-cubic-to-arbitrary}
\end{figure}

We now want to generalize our result Eq.\eqref{eq:bispectrum-squeezed-cubic-interactions} for cubic interactions to arbitrary order self-interactions, as depicted on~Fig.~\ref{fig:from-cubic-to-arbitrary}:
\begin{equation}
	\mathcal{L}_{\sigma}^{\text{int}} = \sum_{i=3} \frac{\lambda_s}{s!} \sigma^s.
	\label{eq:sigma-interaction-terms-arbitrary-nd}
\end{equation}
As before, we will first compute correlation functions with $N$ super-horizon modes before performing the long-short mode split and integrating over super-horizon modes.

When thinking about the possible diagrams that can be constructed when adding super-horizon modes, we see that those modes can either be added to the three sub-horizon modes or to the central vertex, as higher order interactions are now allowed.  We therefore split the $N$ super-horizon modes in four sets (some of those sets can be empty) containing $n, m, \ell$ and $\ell^{\prime}$ long modes that will connect to $p_1,p_2, k$ and to the central vertex respectively, with $N=n+m+\ell+\ell^{\prime}$. 

Owing to the higher-order interaction terms of the hidden sector field, there are more ways to attach the super-horizon modes onto their sub-horizon mode. Therefore, the set of integrals that were put to the power $n,m$ or $\ell$ in Eq.\eqref{eq:correlation-fct-N3} now becomes
\begin{equation}
 \left[ \int d\tau a^4 \lambda_3 \sigma(p_1) \sigma(p_1) \sigma(q) \right]^{n} \to p(n) \hspace{-1em} \sum_{ \left\{n_i | \sum i n_i =n \right\} } \prod_{i=1}^n \left[ \int d\tau a^4 \lambda_{i+2} \sigma(p_1) \sigma(p_1) \sigma^i(q) \right]^{n_i},
\end{equation}
where the sum is performed over the partition of $n$ in $n_i$'s with $\sum i \, n_i =n$. The function $p(n)$ is the partition of $n$ and corresponds to the number of ways to arrange $n$ long modes into $n_i$ vertices with $i+2$ legs. Each ``leg'' integral is then evaluated as
\begin{equation}
	\int d\tau a^4 \lambda_{i+2} \, \sigma(p) \sigma(p) \sigma^i(q) \sim \lambda_{i+2} \ H^{i-2}\  \frac{p^{(\nu-\tfrac{3}{2})i}}{q^{i\nu}}.
\end{equation}
The central vertex integral is:
\begin{equation}
	\int d\tau a^4 \lambda_{i+3} \sigma(p_1) \sigma(p_2) \sigma(k) \sigma^i(q) \sim \lambda_{i+3} H^{i-1} \frac{(k+p_1+p_2)^{\nu(i+3)-\tfrac{3}{2}(i+1)}}{(p_1p_2kq^i)^{\nu}}.
\end{equation}
Collecting all the terms, we get
\begin{equation}
\begin{split}
	F_{N+3}(p_1,p_2,k,q_1,\dots,q_N) & \propto \left( \frac{\Delta_{\zeta} \rho}{H} \right)^{N+3} H^N \frac{\left( p_1 + p_2 + k\right)^{3\nu-\tfrac{3}{2}}}{(p_1p_2kq^N)^{3/2+\nu}} \\
	&\quad \times \hspace{-1em} \sum_{\substack{\left\{n,m,\ell,\ell^{\prime} \, | \right. \\ \left. n+m+\ell+\ell^{\prime}=N\right\}}} \hspace{-1em} \lambda^{\text{eff}}_{n,m,\ell,\ell^{\prime}} \frac{(p_1+p_2+k)^{(\nu-3/2)\ell^{\prime}}}{(p_1^n p_2^m k^{\ell})^{\tfrac{3}{2}-\nu}},
	\label{eq:FN3-arbitrary}
\end{split}
\end{equation}
where again, the sum is over the partition of $N$ in four sets with $n+m+\ell+\ell^{\prime}=N$, and where $\lambda^{\text{eff}}_{n,m,\ell}$ is defined by:
\begin{equation}
\begin{split}
 	\lambda^{\text{eff}}_{n,m,\ell,\ell^{\prime}} \equiv & \frac{\lambda_{\ell^{\prime}+3}}{H} \, p(n) \, p(m) \, p(\ell) \left[ \sum_{\{ n_i | \sum i n_i = n \}} \prod_{i=1}^n \left(\frac{\lambda_{i+2}}{H^2}\right)^{n_i}  \right] \\
	& \times \left[ \sum_{\{ m_i | \sum i m_i = m \}} \prod_{i=1}^m \left(\frac{\lambda_{i+2}}{H^2}\right)^{m_i}\right] \left[ \sum_{\{ \ell_i | \sum i \ell_i = \ell \}} \prod_{i=1}^{\ell} \left(\frac{\lambda_{i+2}}{H^2}\right)^{\ell_i} \right].
\end{split}
\end{equation}

Eq.\eqref{eq:FN3-arbitrary} shares a lot of structure with~Eq.\eqref{eq:FN3-cubic}, and therefore, the derivation of the observed kernel in the sub-volume $N_2^{\text{obs}}(\mathbf{p}_1,\mathbf{p}_2,\mathbf{k})$ remains similar to what was done in the previous section. The result is:
\begin{equation}
\begin{split}
	N_2^{\text{obs}}(\mathbf{p}_1,\mathbf{p}_2,\mathbf{k}) & \sim \left(\frac{\rho}{H}\right)^3 \frac{(2\pi^2)^2}{\Delta_{\zeta}} \frac{(p_1+p_2+k)^{3\nu-\tfrac{3}{2}}}{(p_1p_2)^{\nu-\tfrac{3}{2}} k^{\nu+\tfrac{3}{2}}} \sum_{N=0} \frac{(N+2)!}{N!2!} \frac{c_{\text{NL}}^{(N+2)}}{\fnl} \\
	& \quad \times \left( \frac{2\pi^2\rho}{\Delta_{\zeta} H} \right)^N \prod_{i=3}^{N+2} \left[ \int_{q_i<k_0} \frac{d^3q_i}{(2\pi)^3} \zeta(q_i) q_i^{\tfrac{3}{2}-\nu} \right] \\
	& \quad \times \hspace{-1em} \sum_{\substack{\left\{n,m,\ell,\ell^{\prime} \, | \right. \\ \left. n+m+\ell+\ell^{\prime}=N\right\}}} \hspace{-1em} \lambda_{n,m,\ell,\ell^{\prime}}^{\text{eff}} \frac{(p_1+p_2+k)^{(\nu-\tfrac{3}{2}){\ell^{\prime}}}}{(p_1^np_2^mk^{\ell})^{\tfrac{3}{2}-\nu}}.
\end{split}
\end{equation}

The equivalent to Eq.\eqref{eq:three-pt-fct-symmetric} is then:
\begin{equation}
\begin{split}
 	\langle \zeta (k_1) \zeta (k_2) Z_2^{\text{obs}} (k_3) \rangle \sim & \frac{\fnl}{\fnl^{\text{obs}}} \left( \frac{\rho}{H} \right)^3 \Delta^3 \frac{(k_1+k_2+k_3)^{3\nu-\tfrac{3}{2}}}{(k_1k_2k_3)^{\tfrac{3}{2}+\nu}} \left[\sum_{N=0} \beta_N \zeta_L^N  \right. \sum_{\substack{\left\{n,m,\ell,\ell^{\prime} \, | \right. \\ \left. n+m+\ell+\ell^{\prime}=N\right\}}} \\
 	& \qquad \left. \times \lambda^{\text{eff}}_{n,m,\ell} \left( \frac{k_1^nk_2^m k_3^{\ell}(k_1+k_2+k_3)^{\ell^{\prime}}}{k_0^N} \right)^{\nu-\tfrac{3}{2}} \right],
 \label{eq:n-pt-fct-symmetric}
 \end{split}
\end{equation}
where 
\begin{equation}
	\beta_N =  \frac{(N+2)!}{N!2!} \frac{c^{(N+2)}_{\text{NL}}}{f_{\text{NL}}} \left( \frac{2\pi^2\rho}{\Delta_{\zeta}} \right)^N.
\end{equation}

Taking one of the mode $k_L$ to be longer than the two others $k_S$ -- but still sub-horizon, the bispectrum reads\footnote{The partition function will enforce the right conditions on the values that $n,m,\ell$ and $\ell^{\prime}$ can take if one restricts the self-interactions terms of $\sigma$, Eq.\eqref{eq:sigma-interaction-terms-arbitrary-nd}, to a single term or a subset of the whole series.}:
\begin{equation}
\begin{split}
	\langle \zeta (k_S) \zeta (k_S) \zeta (k_L) \rangle & \sim \frac{\fnl}{\fnl^{\text{obs}}} \left( \frac{\rho}{H} \right)^3 \frac{(2\pi^2)^2 2^{3\nu-\tfrac{3}{2}}}{\Delta_{\zeta}} P(k_L) P(k_S) \left(\frac{k_L}{k_S} \right)^{\tfrac{3}{2}-\nu} \\
	& \quad \times \sum_{N=0} \beta_N \zeta^N_L \left( \frac{k_S}{k_0} \right)^{N(\nu-\tfrac{3}{2})} \sum_{n=0}^{N} \left( \frac{k_L}{k_S} \right)^{n(\nu-\tfrac{3}{2})} \hspace{-2em} \sum_{\substack{\left\{m,\ell,\ell^{\prime} \, | \right. \\ \left. n+m+\ell+\ell^{\prime}=N\right\}}} \hspace{-1em}  \lambda_{n,m,\ell,\ell^{\prime}}^{\text{eff}}.
	\label{eq:bispectrum-squeezed-n-interactions}
\end{split}
\end{equation}
What we observe is that the scaling of Eq.\eqref{eq:bispectrum-squeezed-n-interactions} is the same as Eq.\eqref{eq:bispectrum-squeezed-cubic-interactions}. This is in agreement with our previous results~\cite{CosmicVariance2015}, where it was observed that for contact-like diagrams, the scaling of the bispectrum in the squeezed limit was unchanged. Here, we see that allowing arbitrary order interactions -- rather than just cubic interactions -- gives rise to a similar correlation function than in the cubic case, but with a different, effective coupling constant.

It therefore seems that allowing more super-horizon modes to connect to an already existing vertex does not influence the shape of the correlation function as sharply as creating a new vertex when adding a super-horizon modes. While the amplitude is affected by the fact that the effective coupling constant is changed, the momentum dependence of the correlation functions remains the same as for the cubic case.

\subsection{Derivative Self-Interactions}

Until now, we have been focusing on how long-short modes coupling through \emph{non-derivative} self-interactions affect the bispectrum. In order to be as generic as possible, we quickly describe how \emph{derivative} interactions may affect the correlation function.

For simplicity, we are going to compare the bispectrum in the squeezed limit in the presence of only one super-horizon mode -- as denoted by the subscript $1H$ -- obtained by two similar interaction Lagrangian:
\begin{equation}
	\mathcal{L}_{4}= \lambda \sigma^4 \qquad \text{vs} \qquad \mathcal{L}_4^{\prime} = \frac{\lambda}{M^2} \, \partial_{\mu} \sigma \, \partial^{\mu} \sigma \, \sigma^2.
\end{equation}
The non-derivative Lagrangian yields the following bispectrum:
\begin{equation}
	\langle \zeta_{k_S} \zeta_{k_S} \zeta_{k_L}\rangle_{1H} \sim \lambda \left( \frac{\rho}{H}\right)^4 \frac{P(k_S) P(k_L) P(k_H)}{\Delta_{\zeta}^2} \left(\frac{k_L}{k_S}\right)^{3/2-\nu} \left(\frac{k_H}{k_S}\right)^{3/2-\nu} 
	\label{eq:quartic-interaction-correlation-function}
\end{equation}
where $k_H$ is the super-horizon mode, while $k_L \ll k_S$ are sub-horizon modes.

For the derivative Lagrangian, in Fourier space, the integral to be evaluated is:
\begin{align}
	\int d\tau \lambda^{\prime} a^4 \partial_{\mu} \sigma \partial^{\mu} \sigma \sigma^2 & \to \int d \tau \frac{\lambda}{M^2} a^4 \left[ \sigma_{k_1}^{\prime} \sigma_{k_2}^{\prime} - \mathbf{k}_1 \cdot \mathbf{k}_2 \, \sigma_{k_1} \sigma_{k_2} \right] \sigma_{k_3} \sigma_{k_4} \nonumber \\
	&\sim \frac{\lambda}{M^2} \frac{(k_1+k_2+k_3+k_4)^{4\nu-1}}{(k_1k_2k_3k_4)^{\nu}} \left[ 1 - \frac{\mathbf{k}_1\cdot \mathbf{k}_2}{(k_1+k_2+k_3+k_4)^2} \right]
\end{align}
where boldface denotes three-vectors. Given the configuration of the three modes in the squeezed limit, where the two short modes are almost anti-parallel to each other, the maximum contribution to the dot product will come from these two short modes, $\mathbf{k}_1\cdot \mathbf{k}_2 \simeq k^2_S$. 

The final correlation function for that vertex is:
\begin{equation}
	\langle \zeta_{k_S} \zeta_{k_S} \zeta_{k_L} \rangle_{1H}^{\text{deriv}} \sim \lambda \left(\frac{\rho}{H}\right)^4 \frac{P(k_S) P(k_L) P(k_H)}{\Delta_{\zeta}^2} \left(\frac{k_L}{k_S}\right)^{3/2-\nu} \left(\frac{k_H}{k_S}\right)^{3/2-\nu} \frac{k_S^2}{M^2}.
	\label{eq:quartic-derivative-interaction-correlation-function}
\end{equation}
This bispectrum is suppressed by a factor of $k_S^2/M^2$ with respect to the non-derivative bispectrum Eq.\eqref{eq:quartic-derivative-interaction-correlation-function}, but scalings are otherwise unaffected. Therefore, it is legitimate to focus on non-derivative self-interactions of the hidden sector field in order to study the effect of the coupling between sub- and super-horizon modes on the squeezed limit of the bispectrum.

\section{How Do Long Modes Affect Correlation Functions?}

In the previous sections, we focused on understanding how super-horizon modes could affect the bispectrum in the squeezed limit. Here, we wish to study how these modes can influence \textit{generic} correlation functions. While giving an exact result would require to perform a lengthy computation for each different correlation function, we here want to bring up some features that can be deducted using similar reasonings as those introduced above.

A first aspect to point out is that, owing to the fact that these super-horizon modes are much longer than the other, sub-horizon modes, they don't significantly affect the momentum structure of the correlation function (i.e. $\mathbf{k}+\mathbf{k}_H \simeq \mathbf{k}$). Therefore, all contributions in integrals involving super-horizon modes can be factored out. This allows us to derive several rules describing how a given correlator is affected by super-horizon modes.

Let's consider a correlation function with $N$ external momenta. There might be several corresponding tree-level diagrams; in particular, the distribution of internal momenta can have its importance. We here assume that the correlation function associated to a diagram with no attached super-horizon modes is known, and we derive how that correlation function will be affected by the super-horizon mode couplings.

One evaluates a diagram's contribution to a correlation function by labeling the external and internal edges, and writing down the integrals corresponding to the interaction vertices. When evaluating vertices integrals, we again assume that the dominant contribution comes from the shortest mode, with $\tau \sim k^{-1}$, as in~\cite{Baumann2011}. In order to keep our notation as generic and symmetric as can be, we write the shortest mode as a sum of all the modes present in the integral. For instance, a cubic vertex with three momenta $k_1,k_2$ and $k_3$ yields:
\begin{equation}
	\int d \tau  \lambda  \frac{H^3}{(-H\tau)^4} \frac{(-\tau)^{-3\nu+\tfrac{9}{2}}}{(k_1 k_2 k_3)^{\nu}} \sim \frac{\lambda}{H} \frac{(k_1+k_2+k_3)^{3\nu-\tfrac{3}{2}}}{(k_1 k_2 k_3)^{\nu}}.
	\label{eq:short-mode-contribution}
\end{equation}

\begin{figure}
\centering%
\begin{subfigure}{.5\textwidth}
	\centering%
	\includegraphics[scale=1]{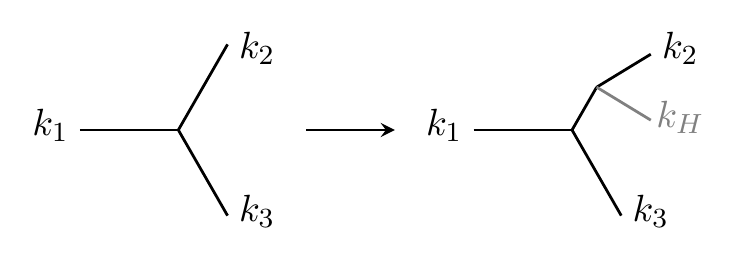}
	\caption{Exchange-like connexion}
	\label{fig:exchange-type}
\end{subfigure}%
\begin{subfigure}{0.5\textwidth}
	\centering%
	\includegraphics[scale=1]{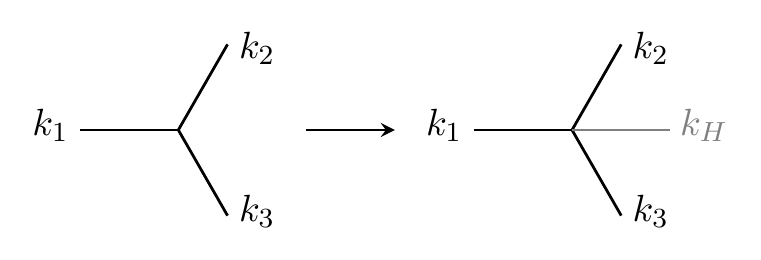}
	\caption{Contact-like connexion}
	\label{fig:contact-type}
\end{subfigure}
\caption{A long mode can be connected to a diagram in two different fashions. It can either be attached to a sub-horizon mode, creating a new vertex (Fig.a), or to an existing vertex (Fig.b).}
\label{fig:types-of-diagrams}
\end{figure}

Given a diagram, there are two ways to attach a super-horizon mode to it: either by connecting the super-horizon mode to a sub-horizon one in an exchange-like fashion, creating a new vertex -- see Fig.\ref{fig:exchange-type}, or by connecting it to an existing vertex in a contact-like way  -- see Fig.\ref{fig:contact-type}. The two cases are treated separately below. In each case, since we have to connect the hidden sector field back to the inflaton sector, there will be a factor of $\rho/H$ and of $\Delta_{\zeta}/k_H^{3/2}$ for each additional super-horizon mode. 

\subsection{Contact-type}
We start with the case of contact-like diagrams, where all super-horizon modes are attached to pre-existing vertices.

Let's first focus on a single vertex, and let's assume it has $m$ sub-horizon modes $p_1, \dots, p_m$
connecting to it. The associated integral will read:
\begin{equation}
	\int d \tau a^4 \lambda_{m} \sigma(\mathbf{p}_1) \dots \sigma(\mathbf{p}_m) \sim \frac{\lambda_m H^{m-4}}{(p_1 \dots p_m)^{\nu}} p_S^{3 - (\tfrac{3}{2}-\nu)m}.
	\label{eq:general-contact-like-no-super-horizon-integral}
\end{equation}
where $p_S$ is the shortest mode of all. If we now also connect $n$ super-horizon modes $q_1, \dots, q_n$ to that same vertex, then the integral becomes
\begin{align}
	\int d\tau a^4 \lambda_{m+n} \sigma(\mathbf{p}_1) \dots \sigma(\mathbf{p}_n) & \sigma(\mathbf{q}_1) \dots \sigma(\mathbf{q}_n) \sim \frac{\lambda_{m+n} H^{m+n-4}}{(p_1 \dots p_m q_1 \dots q_n)^{\nu}} p_S^{3-(\tth-\nu)(m+n)} \nonumber \\
	& \sim  \frac{\lambda_m H^{m-4}}{(p_1 \dots p_m)^{\nu}} p_S^{3 - (\tfrac{3}{2}-\nu)m} \times \frac{\lambda_{m+n}}{\lambda_m} \frac{H^n}{(q_1 \dots q_n)^{\nu}} p_S^{(\tth-\nu)n}
\end{align}
where in the last line, we explicitly separated the part ot the result looking like the integral with no super-horizon modes~Eq.\eqref{eq:general-contact-like-no-super-horizon-integral}, and the super-horizon modes contribution.

If we now consider the more general case where a diagram has $M$ external sub-horizon modes $k_1,\dots, k_M$, $N$ external super-horizon modes $q_1, \dots,q_N$ and $V$ vertices, then the associated correlation function will read:
\begin{align}
\langle \zeta(\mathbf{k}_1) \dots \zeta(\mathbf{k}_M) \rangle_{N} \sim \frac{(\Delta_{\zeta} \rhoh)^{M+N}}{(k_1 \dots k_M q_1 \dots q_M)^{3/2}} \prod_{i=1}^{V} \frac{\lambda_{M_i+N_i}H^{M_i+N_i-4}}{(p_1\dots p_{M_i} q^{N_i})^{\nu}} p_{S_i}^{3-(\tth-\nu)(M_i+N_i)},
\label{eq:general-contact-like-zetaM-pre}
\end{align}
where $p_1,\dots,p_{M_i}$ denotes the number of sub-horizon modes (not necessarily external edges) connected to the vertex $V_i$, $p_{S_i}$ is the shortest of these modes, and $N_i$ is the number of super-horizon modes connected to the vertex $V_i$ such that $\sum_i N_i = N$.

If we gather together the contributions from the super-horizon modes, we can write~Eq.\eqref{eq:general-contact-like-zetaM-pre} as:
\begin{equation}
	\langle \zeta(\mathbf{k}_1) \dots \zeta(\mathbf{k}_M) \rangle_{N} \sim \langle \zeta(\mathbf{k}_1) \dots \zeta(\mathbf{k}_M) \rangle_{0}  \prod_{i=1}^{V} \frac{\lambda_{M_i+N_i}}{\lambda_{M_i}} \left(\frac{ \rho \Delta_{\zeta}}{q^3}\right)^{N_i} \left(\frac{q}{p_{S_i}}\right)^{N_i (\tth - \nu)} \hspace{-1em}
\end{equation}
where the subscript $0$ denotes the correlation function with zero super-horizon modes. This result is intermediate, as super-horizon modes need to be integrated over, but we can already notice that there is an additional momentum dependence $p_{S_i}^{N_i(\tth-\nu)}$, which shows that the correlation function will only be affected by the shortest mode of each vertex.

While performing the integration over the super-horizon modes is beyond the scope of this work, we give a postulate of the final form of expression~\eqref{eq:contact-type-final-correlation-function-postulat}. First, we notice that the correction due to super-horizon modes is an overall multiplicative factor; we also note that the super-horizon momentum distribution is of the correct form to give a factor of $\zeta_L$ when integrated over, as in~\cite{CosmicVariance2015}; and finally, there are many similarities between Eq.\eqref{eq:contact-type-final-correlation-function-postulat} and the corresponding result for the bispectrum, which suggests that the procedure of integrating over the super-horizon modes can be carried out in a similar fashion.
Therefore, our postulate for the final expression of the locally observed correlation function is:
\begin{equation}
	\langle \zeta(\mathbf{k}_1) \dots \zeta(\mathbf{k}_M) \rangle^{\text{obs}} \sim \langle \zeta(\mathbf{k}_1) \dots \zeta(\mathbf{k}_M) \rangle_{0} \times \sum_{N=0}  \zeta_L^N \prod_{i=1}^{V} c_{\text{NL}}^{(N,i)}  \left(\frac{p_{S_i}}{k_0} \right)^{(\nu-\tth)N_i}.
	\label{eq:contact-type-final-correlation-function-postulat}
\end{equation}
The result for the particular case of the squeezed bispectrum, presented in~\cite{CosmicVariance2015}, Eq.(3.43), is recovered by the expression~\eqref{eq:contact-type-final-correlation-function-postulat}.

\subsection{Exchange-type}

We now turn to the case where super-horizon modes are attached to a sub-horizon modes by creating a new vertex. Let's consider a diagram with $m$ external edges, whose corresponding correlation function can be written as:
\begin{equation}
	\langle \zeta(k_1) \dots \zeta(k_m) \rangle = \frac{\Delta^{m}}{(k_1\dots k_m)^{\tfrac{3}{2}}} \mathcal{F}^{\prime}(k_1,\dots,k_m).
	\label{eq:correlation_function_generic_0H_exchange}
\end{equation}

If we now connect a super-horizon mode $q$ to one of the main, sub-horizon mode $k$ (which can be either external or internal) of the diagram, this creates a new vertex:
\begin{equation}
	\int d \tau a^4 \lambda_{3} \, \sigma(k) \sigma(k) \sigma(q) \sim  \frac{\lambda_3}{H} \frac{k^{\nu-\tfrac{3}{2}}}{q^{\nu}}.
	\label{eq:cubic-vertex-exchange}
\end{equation}
Therefore, the new correlation function with one super-horizon mode connected to it becomes:
\begin{align}
	\langle \zeta(k_1) \dots \zeta(k_m) \rangle_{1} & = \frac{\Delta^{m}}{(k_1\dots k_m)^{\tfrac{3}{2}}} \mathcal{F}^{\prime}(k_1,\dots,k_m) \frac{2\pi^2\lambda_3}{\Delta_{\zeta}H} P(q) \left(\frac{k}{q}\right)^{\nu-\tfrac{3}{2}}\\
	& = \frac{2\pi^2\lambda_3}{\Delta_{\zeta}H} P(q) \left(\frac{q}{k}\right)^{\tfrac{3}{2}-\nu} \langle \zeta(k_1) \dots \zeta(k_m) \rangle_0,
	\label{eq:correlation_function_generic_1H_exchange}
\end{align}
where $k=k(k_1,\dots,k_m)$ is the sub-horizon mode on which the super-horizon modes has been attached.

Similarly, if $N$ super-horizon modes are connected to -- internal or external -- edges of the diagram, the correlation function changes:
\begin{equation}
	\langle \zeta(k_1) \dots \zeta(k_m) \rangle_{N} = \left(\frac{2\pi^2\lambda_3}{\Delta_{\zeta}H} P(q)\right)^N \left( \prod_{i=1}^{N} \left(\frac{q}{k_{i}}\right)^{\tfrac{3}{2}-\nu} \right) \langle \zeta(k_1) \dots \zeta(k_m) \rangle_0,
	\label{eq:correlation_function_generic_NH_exchange}
\end{equation}
where $k_{i} = k_i(k_1,\dots,k_m)$ is the sub-horizon mode to which the super-horizon mode $q$ has been connected.

In this case, our ansatz for the locally observed correlation function where all super-horizon modes have been connected by creating a new vertex is:
\begin{equation}
	\langle \zeta(k_1) \dots \zeta(k_m) \rangle^{\text{obs}} = \langle \zeta(k_1) \dots \zeta(k_m) \rangle_0 \times \sum_{N=0} c^{\prime (N)}_{\text{NL}} \zeta_L^N \prod_{i=1}^N \left(\frac{k_i}{k_0}\right)^{\nu-\tth}.
	\label{eq:correlation-function-ansatz-exchange}
\end{equation}
We notice that the momentum dependence of Eq.\eqref{eq:correlation-function-ansatz-exchange} is different than the result from the previous section Eq.\eqref{eq:contact-type-final-correlation-function-postulat}. For contact-like connexion, the momentum dependence is affected by the vertex's smallest mode in presence, while for exchange-like connexion, the momentum involved is the momentum of the mode onto which the super-horizon mode connects.

\section{Discussion}

The presence of additional fields during the inflationary epoch can have a significant impact on the statistics observed in a Hubble, sub-volume by a local observer. Hidden sector fields are not restricted by the inflation's (approximate) shift symmetry, and can hence have a variety of self-interactions. They also enable couplings between super-horizon and sub-horizon modes, inducing a cosmic bias on the observed correlation functions. It is therefore important to understand how these coupling can affect the locally observed cosmological parameters.

We work in the context of Quasi-Single Field Inflation where a single hidden sector field is allowed to have arbitrary order non-derivative self-interactions. We have shown how the squeezed limit of the bispectrum is affected by couplings of super-horizon modes with observable modes. 
The influence of super-horizon modes can be separated in several categories, depending on the way they couple to the sub-horizon fields. It was shown previously that, when connected to a preexisting vertex, super-horizon modes change the coupling constant of the vertex, giving rise to an effective coupling constant, but don't affect the scaling between long and short modes.
However, we see that when connected to a sub-horizon mode by creating a new vertex, the scaling of $k_L/k_S$ is changed. This gives rise to an unavoidable cosmic variance, which prevents us from determining the spectrum of masses of the hidden sector field with unlimited accuracy. We gave a quantitative estimation of the amplitude of the bias on the mass of the hidden sector field, which can amount to a 1-$\sigma$ uncertainty of 10\%; this results depends on the parameters of the model, which have been chosen to ensure that the system is weakly non-Gaussian.

For completeness, we also gave a short comparison between the effect of non-derivative and derivative interactions. Derivative interactions are suppressed with respect to non-derivative ones, which is the reason why we focused on non-derivative interactions.

Finally, the domain consequences of super-horizon modes coupling is not limited to the power spectrum or the bispectrum. We hence described how arbitrary order correlation functions are affected by unobservable modes, depending again on how they connect to a specific diagram. We show that some features -- such as the scaling of the different modes involved in the correlator, or the rise of an effective coupling constant  -- can be easily estimated by using our method. We hope this will can be put in application in the future to understand how other cosmological parameters are affected by unobservable modes.

\paragraph{Acknowledgment:} We thank Sarah Shandera for helpful discussions and the numerous comments on the draft. We thank the Perimeter Institute for hospitality where this work was partially conducted. This research was supported in part by Perimeter Institute for Theoretical Physics. Research at Perimeter Institute is supported by the Government of Canada through Industry Canada and by the Province of Ontario through the Ministry of Economic Development \& Innovation.

\end{document}